\documentclass[10 pt, conference]{IEEEtran} 
\IEEEoverridecommandlockouts

\usepackage[utf8]{inputenc}
\usepackage[T1]{fontenc}
\usepackage[inline]{enumitem}
\usepackage[numbers]{natbib}

\title{
Automatically Identifying Parameter Constraints in Complex Web APIs: A Case Study at Adyen
}

\author{\IEEEauthorblockN{Henk Grent, Aleksei Akimov}
\IEEEauthorblockA{\textit{Adyen N.V.} \\
Amsterdam, the Netherlands\\
\{Henk.Grent,Aleksei.Akimov\}@adyen.com}
\and
\IEEEauthorblockN{Maurício Aniche}
\IEEEauthorblockA{\textit{Delft University of Technology}\\
Delft, the Netherlands \\
M.F.Aniche@tudelft.nl}
}

\usepackage{framed} 
\usepackage{url}
\usepackage{booktabs}
\usepackage{graphicx}
\usepackage{algorithm2e}

\usepackage[pdftex,dvipsnames]{xcolor}

\usepackage{listings}
\usepackage{xcolor}

\usepackage{color}
\definecolor{dkgreen}{HTML}{72A98F}
\definecolor{mauve}{HTML}{834D93}
\definecolor{midnightblue}{HTML}{290DFD}

\newcommand{\rqone}{How effective are documentation- and static code analysis in identifying parameter constraints in a large-scale enterprise API?}

\newcommand{\rqtwo}{What are challenges faced by documentation and code analysis techniques that identify inter-parameter constraints in a large-scale enterprise API?}

\expandafter\def\expandafter\UrlBreaks\expandafter{\UrlBreaks
  \do\a\do\b\do\c\do\d\do\e\do\f\do\g\do\h\do\i\do\j%
  \do\k\do\l\do\m\do\n\do\o\do\p\do\q\do\r\do\s\do\t%
  \do\u\do\v\do\w\do\x\do\y\do\z\do\A\do\B\do\C\do\D%
  \do\E\do\F\do\G\do\H\do\I\do\J\do\K\do\L\do\M\do\N%
  \do\O\do\P\do\Q\do\R\do\S\do\T\do\U\do\V\do\W\do\X%
  \do\Y\do\Z}
  
\begin{document}

\maketitle
\thispagestyle{empty}
\pagestyle{empty}

\begin{abstract}

Web APIs may have constraints on parameters, such that not all parameters are either always required or always optional. Moreover, the presence or value of one parameter could cause another parameter to be required, or parameters could have restrictions on what kinds of values are valid. Having a clear overview of the constraints helps API consumers to integrate without the need for additional support and with fewer integration faults.

We made use of existing documentation and code analysis approaches for identifying parameter constraints in complex web APIs. 
In this paper, we report our case study of several APIs at Adyen, a large-scale payment company that offers complex Web APIs to its customers. Our results show that the documentation- and code-based approach can identify 23\% and 53\% of the constraints respectively and, when combined, 68\% of them. 
We also reflect on the current challenges that these approaches face. In particular, the absence of information that explicitly describes the constraints in the documentation (in the documentation analysis), and the engineering of a sound static code analyser that is sensitive to data-flow, maintains longer parameter references throughout the API's code, and that is able to symbolically execute the several libraries and frameworks used by the API (in the static analysis).
\end{abstract}

\begin{IEEEkeywords}
software engineering, web APIs, parameter constraints inference.
\end{IEEEkeywords}

\section{Introduction}
\label{cha:intro}

Web Application Programming Interfaces (Web APIs) allow applications to access the functionality or data of a service through HTTP requests. Web APIs commonly provide an API reference \cite{maalej2013patterns}, which describes what operations are available through which endpoint and which parameters are required or optional for requests to these endpoints. However, these parameters are not always just required or optional: whether they are required can depend on the presence or value of another parameter \cite{oostvogels2017inter, martin2019catalogue}. 

Within Adyen, as a payment platform, we observe a large number of such inter-parameter constraints. Take constraints that apply on different payment methods as an example; if one of our API consumers want to make a payment with iDEAL, then the previously non-required \textit{issuer} and \textit{returnURL} parameter are now required. For other payment methods, different parameters become required. As another example, when authorising a payment, the API expects a bank account or a card as payment details. Without either the request will fail.

Having a clear overview of the constraints in a Web API is highly important in practice, as it helps API consumers to integrate with our API without the need for company support. Incomplete or incorrect documentation on these constraints can waste a lot of time, and cause costly integration faults, as we have observed before~\cite{aue2018exploratory}. Currently, these constraints are documented and maintained manually by the API developers, which can be a laborious and difficult task. This difficulty comes from the size and complexity of the code base of the web service, and documentation being provided by different people than those who write the code. Therefore, tools that help API developers identify and maintain the constraints in their APIs are needed.

In this paper, we report our case study on applying existing approaches in the literature to automatically identify constraints in our large-scale complex Web APIs. For the main APIs under study, the Adyen APIs, complexity largely results from making a large number of payment-related operations available through a single interface. Adyen's API contains several endpoints, with a varying number of parameters\footnote{See \url{https://docs.adyen.com} for a complete picture of the APIs we provide.}. For example, version 52 of the ``/payments'' endpoint features 55 top-level parameters and 371 parameters in total. 

The approach we implemented makes use of two different sources: the online service documentation we provide to our customers, and the source code of our API. We draw inspiration from Wu et al. \cite{wu2013inferring} who set out to identify inter-parameter constraints from the online API reference and available software development kits (SDKs). 

When compared to related work, we anticipate two challenges. First, the complexity of our API. As mentioned before, our APIs have a significantly higher number of parameters and inter-dependencies. It is not clear whether existing approaches scale to this size. Second, the complexity of our codebase. Our code base makes use of different frameworks and abstractions, and the business rules executed by a single API call may be spread across several classes. Therefore, it is not clear whether the proposed program analysis techniques will effectively be able to extract the required information out of the source code.

Our results show that the documentation and code-based approaches can identify 23\% and 53\% percent of the inter-parameter constraints, respectively. When the constraints identified by both approaches are combined, a total of 68\% of the inter-parameter constraints can be identified. Moreover, the code analysis is able to identify 78\% of the single-parameter constraints. 

We observe that the two approaches face largely separate challenges. The documentation based approach suffers from a lack of available explicit information describing the constraints. The static code analysis tends to be able to extract constraints from the source code by maintaining a basic variable stack, evaluating method calls, and analysing conditions in for-loops, switch and if-else statements. However, it faces challenges related to the engineering of a sound static code analyser that is sensitive to data-flow, maintains longer parameter references throughout the API's code, and that is able to symbolically execute the several libraries and frameworks used by the API.


The main contributions of this paper are:

\begin{itemize}
  \item An empirical study demonstrating the effectiveness of code-based and documentation-based inter-parameter constraints identification approaches in a large-scale complex Web API.
  
 \item A set of challenges that existing code-based and documentation-based approaches face when analysing large-scale complex Web APIs.
\end{itemize}

\section{Related Work}
\label{cha:related}

\subsection{API Usability and Constraints in Practice}

The literature on API usability has been increasingly growing, as shown in \citet{rauf2019systematic}'s literature review.
Usability, as an aspect of software quality, frequently takes focus in API design literature as, ultimately, APIs are consumed by people to create specific functionality for their own use case. 

The lack of documentation is a key obstacle for API learnability \cite{robillard2011field, piccioni2013empirical}. \citet{robillard2011field} identify five documentation factors impacting the developers learning experience: documentation of intent, code examples, mapping usage scenarios, penetrability, and format and presentation.
For developers, API usability is key in the adoption/integration process. Learning obstacles may result in opting for a different service \cite{rauf2016perceived, robillard2011field} or increased integration efforts in supporting API consumers. Research suggests that a significant portion of faults in API integration can be attributed to invalid or missing user input \cite{aue2018exploratory}. These integration faults relate to parameter constraints, such as the absence of (conditionally) required parameters or invalid values for provided parameters.

Oostvogels \citep{oostvogels2017inter} describes three categories of constraints in APIs:
\begin{enumerate*}[label=(\roman*)]
\item exactly one of a set of parameters must be present,
\item the presence or value of one parameter depends on the presence or value of another parameter,
\item a group of parameters should either all be present or not present.
\end{enumerate*}
These three categories are types of \emph{inter-parameter constraints}, as they describe a requirement on the presence or value of a parameter based on the presence or value of another parameter (e.g., if ``country'' is NL, then ``payment method'' should be ``iDeal''). In addition, in this paper we also study \emph{single-parameter constraints}, which describe the requirement on the presence or value of a single parameter (e.g., country should be either ``NL'' or ``BE'').

The work by Martin-Lopez et al. \cite{martin2019catalogue} gives an overview of the frequency of inter-parameter constraints for different industries, considering REST APIs. According to their work, 85\% of the REST APIs have inter-parameter constraints and on average 9.8\% of the operations contain constraints. Moreover, most of the constraints in the wild are not complex, and only 4\% of the dependencies in REST APIs are classified as complex \cite{martin2019catalogue}. In less expansive studies, Oostvogels \cite{oostvogels2017inter} and Wu et al. \cite{wu2013inferring} report comparable numbers.


\subsection{Automatic Inference of Parameter Contraints}

A handful of papers exist outlining approaches for automatically identifying single and inter-parameter constraints. These approaches rely on documentation, API responses, or code analysis to infer such constraints for simple APIs.

Gao et al. \cite{gao2014inferring} uses a decision tree based approach to infer inter-parameter constraints. The information for populating the decision tree is inferred from observing API responses for a given candidate constraint. These candidates are chosen using a set of heuristics and by observing the API's feedback. The latter includes parsing error messages provided as feedback by the API. While the approach was able to infer 145 out of 154 manually identified constraints, just a few APIs were evaluated. The APIs under study did contain at around five parameters per endpoint on average.

Pandita et al. \cite{pandita2012inferring} utilize a number of sources of documentation, including in-code comments, to infer constraints using a NLP based pipeline. These constraints are both inter-parameter constraints as well as single parameter constraints. A large part of the pipeline is responsible for transforming natural text to formal contracts. The constraints are not automatically validated for correctness. This approach yields an average of 92\% precision and 93\% recall on a number of Facebook Web APIs and .NET libraries.

The work by Atlidakis et al. \citep{atlidakis2018rest} use a fuzzing type approach to find dependencies between parameters for different endpoints. That is, it aims at identifying dependencies between a parameter in endpoint A and another endpoint B. To steer the fuzzing process, OpenAPI specifications and the feedback from API responses are used. The fuzzing approach fires a larger number of API requests, at around 5000.

These approaches are designed to infer constraints from documentation, but do not consider code as input. 
Wu et al. \cite{wu2013inferring} set out to automatically identify inter-parameter constraints by inferring constraint candidates from the online API reference and available software development kits (SDKs). These candidates are then verified by calling the public web service with request bodies which would satisfy or violate the candidate constraints.
Their approach uses a combination of NLP and data flow analysis, for documentation and SDK analysis respectively. 
The approach achieves a precision and recall of around 95\% on four (less complex) APIs. Although the approach relies on both code and documentation by design, the results indicate that by far most of the inter-parameter constraints are inferred from the documentation. The documentation provided a total of 351 candidates and the code a total of 36 candidates. The documentation based candidates did have a lower precision than the SDK based candidates, at 20.8\% and 100.0\% respectively, but opposite being true for recall at 82.9\% and 40.9\% respectively.
We use Wu et al.'s~\cite{wu2013inferring} approach as inspiration for our approach.

\section{Approach}
\label{cha:approach}

\begin{figure}
  \begin{center}
    \includegraphics[width=\linewidth]{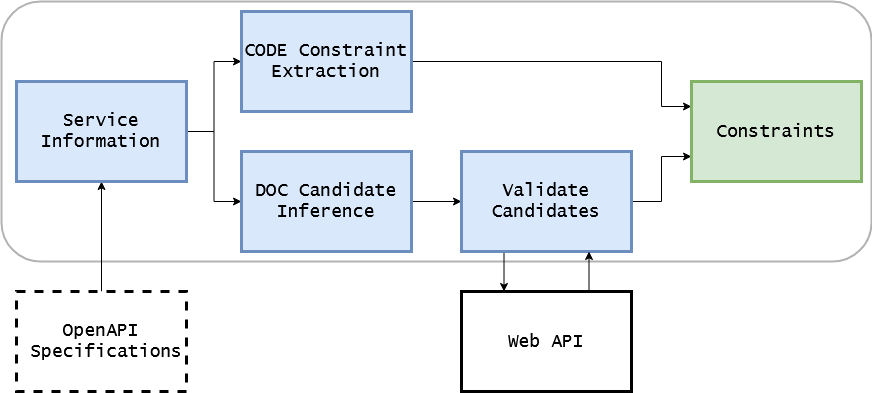}
  \end{center}

  \caption{Overview of the process, showing the initial service information collection step and then the documentation- and code analysis approach.}
  \label{fig:architecture}
\end{figure}

We show a high-level overview of our approach, inspired by Wu et al. \cite{wu2013inferring}, in Figure \ref{fig:architecture}. We shortly describe the general process, and later describe the documentation and code analysis in more detail.

In the first step, we collect information about the parameters for the endpoints of a Web API from the OpenAPI Specification (OAS)\footnote{The OAS is an API description standard which provides service information in a structured way, typically using the JSON format.}. More specifically, we extract the data type of each parameter, whether the parameter is required, any enum values, their description, and parent- and sub-parameters. This information aids us with several tasks, such as default value generation for making requests to an API and detecting parameter references. We use this information in both approaches.

For the documentation based approach, we analyze the textual documentation to infer constraints. This process has two steps. First, we extract sets of candidate parameters from the OAS descriptions. We explain how we obtain these candidates in Section \ref{sec:doc_analysis}.
Secondly, we validate the candidates that were collected in the previous step. We rely on sending requests to the API to infer the constraint. Whether a request fails or succeeds tells us whether it satisfied the requirements of an API or not.

For the code analysis approach, we look at the control structure of methods within the source code to extract constraints. We aim to infer the usage of parameters within this control structure and the preconditions that apply to their use. For example, $if(X \; != null) \{Y\}$ would allow us to infer that $Y$ is needed with the precondition that $X$ is provided in the request. 

\subsection{Documentation Analysis}\label{sec:doc_analysis}
With documentation analysis, we aim to infer if there are constraints between parameters by analyzing the textual documentation of the Web API. Documentation analysis has two distinct steps: finding sets of parameters which might have constraints between them (candidates) and then determining the exact inter-parameter constraints by means of sending API requests to the subject API (validation).

\subsubsection{Candidate Inference}
We use the OAS' parameter descriptions to find candidates. Adyen's API Explorer\footnote{\url{https://docs.adyen.com/api-explorer/#/PaymentSetupAndVerificationService/v52/post/payments}} visualizes these parameters, along with their descriptions, for all public endpoints. The intuition is that the description of one parameter can refer to another parameter, which hints at a possible constraint between the two parameters, e.g., given the parameter \textit{bankAccount} with the description ``The details of the bank account. Either \textit{bankAccount} or \textit{card} is required.'', we assume the two can have a constraint between them.

To extract candidates, we use a co-occurrence matrix \cite{bordag2008comparison}. This co-occurrence matrix contains a row and column for every parameter in a given endpoint. To populate this co-occurrence matrix, we automatically analyze the description of every parameter; if the description of a parameter contains the name of another parameter, then their corresponding entry in the matrix is updated, e.g., for the earlier example the cell corresponding with \textit{bankAccount} and \textit{card} will be updated by one.

Whether a parameter is required may depend on the value of another parameter, e.g., $paymentMethod \:=\: "iDEAL" \rightarrow returnUrl$. Such value-dependent constraints require additional information to be inferred in the validation step. More specifically, we need to know which values are relevant for what parameters. We do this by checking if the descriptions mention any of the enum values the OAS provides.
When an enum value of a parameter is mentioned in a description, then this value is marked and used in the subsequent validation step.

Certain parameters may occur extraordinarily often in descriptions. This is often because of parameter names being common as a word in natural text. Words such as 'reference' and 'value' tend to be used without it being a reference to a parameter. This would yield us a lot of irrelevant candidates. Consequently, we ignore parameters that co-occur with too many other parameters.\footnote{We experimented with different values to determine what ``too many'' would mean. At the end, we decided to ignore parameters that co-occurred more than twice the average.}

\subsubsection{Validating Candidates}
From the documentation analysis, we get sets of parameters which might have constraints between them (candidates), and for each parameter which values were found in the documentation. The aim is to figure out the exact inter-parameter constraint that applies to these parameters, if any. We do this by generating requests and observing the API response for failure. If a request fails, this tells us that some constraint was not satisfied.

We generate a table for each candidate. In this table, each row indicates the present, or absent parameters and whether the corresponding request's result was successful. We represent all the possible combinations of parameters in such a table. For each row in this table, a base request is generated with the parameters indicated as present included and the parameters indicated as not present removed. This request is then sent to the API, and the response is checked for failure. If the request fails, then the Result column is updated accordingly. 

Building valid requests is a major part of validating the previously generated candidates. To that aim, we build request bodies by modifying the \textit{base request} according to the modifications imposed by the table. Following such a table, the parameters indicated as present are included and the parameters indicated as not present removed from the base request. 

The base request is a default request specified for each endpoint, which should always succeed. These default requests can either be specified manually, or they can be generated from the OAS. Including all parameters specified as required by the OAS typically results in a valid base request. When this was not the case, we manually added the missing parameters to the base request.

The parameters provided in a request need to have valid values. What qualifies as valid depends on what values are meaningful for the given parameter. For example, if a parameter represents a date providing any value which is not a date makes little sense. We use either a manually defined value or default value. The default value depends on the type of the parameter.

The type of the parameter can be inferred from the OAS, which are currently defined as \textit{string}, \textit{number}, \textit{integer}, \textit{boolean}, \textit{array} and \textit{object}\footnote{\url{https://swagger.io/docs/specification/data-models/data-types/}}. For each of these types a standard value can be configured. A string may by default return 'str' and an integer may return '0'. For some parameters, such a default value may not be sufficient, i.e., the request always fails with the default value. In such cases, one has to manually define a standard value. This often applied to parameters such as card numbers and account names.

Whether a request was successful or not is determined primarily by the HTTP status code returned as a response to a request. Generally, a 2xx is considered as a success, and a 4xx or a 5xx as a failure. 

\subsection{Code Analysis}
With code analysis, we aim to extract constraints from the control structure of the source code. For this, we analyze methods relevant to handling the HTTP requests made to the API. A method called the 'controller method' is typically responsible for handling requests made to one endpoint of an API. Starting from the controller method, we detect the access of parameters and analyze any control structures and method calls parameters are used in. Within our case study, the Web APIs primarily use Java. As such, the control structures mostly include if-else statements, switch-statements, and for-loops.

Following the code snippet in Figure~\ref{alg:example_bank_card}, we can see how we could infer the dependency of the \emph{card} on \emph{bankAccount} and the constraint on the value of \emph{offset}. That is, if the \emph{card} is not provided for payment details, then we would need the \emph{bankAccount} from the request. For the \emph{offset}, we know that it should be smaller or equal to 80. In practice, there is a large number of challenges involved in inferring such dependencies, this example establishes a basic intuition.

\begin{figure}
\centering
\begin{lstlisting}[language=Java]
void handle(Request req){   
  if(req.getCard() != null) {
      method = req.getCard()
      validateCard(method)
  } else {
     method = req.getBankAccount()
  }
  if(req.getOffset() > 80) {
      throw Exception()
  }
  ...
  method.preprocess()
}
\end{lstlisting}
\caption{An example method handling an API request.}
 \label{alg:example_bank_card}
\end{figure}

\subsubsection{Control Flow Graph}
\label{sec:control_flow_graph}

To represent the control structure of a method, we use a control-flow graph (CFG). We generate such CFGs for every method we analyze. The CFG shows what branches can be taken, and as such it can be used to know what parameters are used within those branches and which preconditions apply for those branches. Since the CFG tells us what branches lead to invalid states, such as throwing exceptions, we can infer what preconditions would cause the request to be invalid. We collect constraints by iterating over the statements of the CFG. In this process, we collect a tree of preconditions and consequences to represent the constraints, e.g., $if(req.getOffset() > 80) \: \{ throw Exception() \}$ gets parsed to $offset > 80 \rightarrow Invalid \: State$.

Looping expressions, such as the for-loop, can be difficult to analyze statically. This is because the condition breaking the loop can be complex. However, we noticed that the exact analysis of looping expressions was not important for the inference of constraints. Looping statements were sometimes used for parameters which have an array value, e.g., "people": $[\{"name":"Frank", ...\}, ...]$. For such array values, any conditions within the body of a loop would apply to all values that would be iterated over. Hence, analyzing the body of a for-loop once was sufficient.

\subsubsection{Sensitivities}
We perform analysis which is flow-sensitive, partially path-sensitive and context-sensitive. In the analysis, the branches of the control flow graph (CFG) are considered without explicitly taking the previously evaluated path into account. This makes the analysis only partially path-sensitive. To exemplify this, consider a node C reachable through either A or B. When evaluating C the program is agnostic to whether the execution trace would have gone through A or B. If a variable is modified in two exclusive branches, then the most recent modification is chosen. Similarly, we do also not keep track of data conditions that would result from taking one path or the other. For example, if a branch has $offset > 80$ as a guard then we do not assume anything about the value of \textit{offset} outside the branch's body.

\subsubsection{Inter-Procedural Analysis} \label{sec:inter_proceudral}
We generate a static call graph, starting from the controller method. We recursively construct this call graph up until a pre-defined depth. For us, a depth of 15 was sufficient.
The bodies of the methods in the call graph are analysed with the arguments that are passed from its calling context. We then recursively merge all the constraints we find in the bodies of the methods throughout the call graph.




\subsubsection{Variable Stack} 
Knowing which variables correspond with which parameter is essential for extracting parameter constraints from the code. When a variable is referenced, we want to know if it is related to a parameter and, as such, relevant for the constraints we will extract. To keep such references, we maintain a variable stack.
The stack keeps track of known concrete values for variables and which parameters correspond with which variable. 

For Java primitives, including strings, we evaluate basic operations such as addition and subtraction. E.g. $"en" + "\_US"$ is resolved as $"en\_US"$. For booleans, we resolve binary operations only if it can be said that they are surely false or true. E.g. given $A || B$ with $A = true$ we know the expression is true. If such expressions are assigned to a variable, then we update the variable stack accordingly. Any expression that we can not resolve result in the value being equal to \emph{null}.

For collections, such as arrays, we keep track of the contents of the collection if the contents are primitive types or enum values. Given APIs often consume simple types, these basic collections are the most significant for inferring constraints. For example, if the stack keeps track of a list of countries in the \textit{countries} variable, and the \textit{country} variable corresponds with the parameter $country$, then later on we could parse the statement $if(countries.contains(country))$ to a meaningful precondition of a constraint. 

\subsubsection{Boolean Function Calls} 
The guards of conditional statements, such as if-statements, may depend on the result of a boolean function call. In order to infer which conditions apply to either a 'true' or 'false' result, we use an adaption to the default approach for analyzing function calls. In this adapted approach, all conditions are collected that would result in the function returning 'true'. For simplicity, we assume functions do not return \textit{null}.

\subsubsection{Common Expressions} 
The core Java language includes common methods whose logic is hard to infer using static analysis, but can still be given meaning to individually due to their common nature. In this case, we do not use the default parsing process, but map the expression to a manually defined machine-readable output. Examples of these are the \textit{.length()} method for strings and the \textit{.equals(arg)} method. We deal with the \textit{.length()} to be able to infer constraints on the length of string type parameters and the \textit{.equals(arg)} operation can be parsed as a simple equality constraint.
We applied the same concept for a handful of common methods used within Adyen.

\subsubsection{Guard Parsing}
We parse guards (i.e., conditional statements such as an \textit{if} statement) as a collection of ANDs and ORs. In the process of parsing these statements, the expressions that occur directly in the guard are evaluated, i.e., any referenced variables are retrieved from the variable stack, and expressions are resolved as described earlier.

\subsubsection{Unparsed Statements}
The parts of conditional statements that can not be parsed to a constraint on a parameter are annotated as \emph{unparsed}, but still shown in the representation. Since code is (often) written to be legible by humans, this allows us to retain some information the condition might have. Suppose the \textit{!isValidCard(card) \& card.getIssuer() != null}. Assume that we could not resolve the reference to \textit{isValidCard(card)}. The guard would be parsed to  \textit{and(!Unparsed(isValidCard(card)), issuer != null)}.

\subsubsection{Duplicated Parameter Names}
In APIs with object encapsulation, the same parameter name may be used multiple times for different parameters. An example of this is the parameter 'reference' in a number of Adyen endpoints. As a result of this, any reference to 'reference' can reference multiple 'reference' parameters.

The correct parameter is inferred from the context of the most recently accessed variables. For example, given we just accessed the 'card' parameter, then we can infer that the 'reference' parameter probably corresponds with 'card.reference' and not (e.g.) 'bank.reference'.

\subsubsection{Request to Object Conversion}


Typically the request passed to an API is deserialized from its original format (JSON, XML) to an object model. Within Adyen parameter names correspond directly with the resulting object fields after deserialization. This allows us to maintain the link between request parameters and fields accessed through the code.


\subsubsection{Identifying Invalid States}
We use throwable exceptions occurring in the code to know if the preconditions leading to that condition should be avoided. Any code statement that tells us the preconditions should be avoided is marked as an 'invalid state'. In most cases, checking the code for such throwable exceptions was enough for extracting constraints. However, there are cases in which parameter constraints may not be enforced by explicit exceptions. Take a try-catch construction in Java, for example. If an error is thrown, we do not directly know what caused it. This may require the use of static null-pointer detection (e.g. \cite{spoto2011precise}). We did not encounter such try-catch constructions; as such, we only dealt with explicit invalid states.

Errors may also be deferred to a later point in program execution. In this case the results of a validation step may be added to a result map, which is later used to throw exceptions. Due to the nature of our static analysis, such flows are difficult to identify. Our solution is to identify patterns, that can be used to identify such a deferred invalid state. For example, any statement containing $x.addError(...)$ could be tagged as an invalid state.

\section{Research Methodology}
To assess the efficiency of documentation-based and code-based inter-parameter constraint identification techniques, we propose the following research questions:

\begin{description}
    \item[RQ$_1$:] \textit{\rqone}
    

    \item[RQ$_2$:] \textit{\rqtwo}
    

\end{description}

In the remainder of this section, we explain the APIs and endpoints we selected, how we build the ground truth used to compare to the output of the approach, and finally how we performed the analysis.

\subsection{Selected Endpoints}
We aimed at selecting a representative set of Adyen APIs and endpoints which were publicly accessible. At the time of writing, there are three distinct public APIs: Checkout, Payments, and Adyen for Platforms. 

We selected endpoints on the basis of the following criteria: an endpoint has to contain inter-parameter constraints, and the internal logic must be dissimilar enough from any previously selected endpoints. This dissimilarity criterion comes from the observation that endpoints frequently featured the same (inter-)parameter constraints, as a result of strong code reuse. Including such similar endpoints would lead to an unbalanced set of endpoints, in which we would effectively be analyzing the same code a number of times.

We selected the following APIs and endpoints: 

\begin{itemize}
    \item Checkout: \textit{/payments}
    \item Payments: \textit{/authorise}, \textit{/capture}, \textit{/storeDetailAndSubmitThirdParty}, \textit{/getCostEstimate}.
    \item Adyen for Platforms: \textit{/createAccountHolder}, \textit{/getAccountHolder}, \textit{/updateAccountHolder}, \textit{/createAccount}, \textit{/uploadDocument}.
\end{itemize}

\subsection{Ground Truth}\label{sec:ground_truth_methodology}
To understand the effectiveness of our approach, we need to compare the obtained results to a known ground truth.
The ground truth used in this study consists out of a representative set of constraints which we manually collected for each of the selected endpoints. These constraints include both inter- and single-parameter constraints.
Three aspects of the ground truth are particularly important: the collection, selection, and representation. Note that we do not publish this ground truth for security reasons.

\subsubsection{Collection}
Given that only a number of constraints were known beforehand, we had to carefully inspect the code of all selected endpoints for constraints. In this process, we start at the controller method and follow the code until its end, taking note of any constraints we find along the way. Any constraints were validated by making API requests corresponding with the constraint in order to ensure their correctness. Additionally, developers from the respective APIs were asked for guidance in pointing out constraints known by them and the general logic of handling requests related to that API.

\subsubsection{Selection}
Some inter-parameter constraints are trivial and, as such, not every constraint which is technically a constraint is included. For example, given we have an \textit{address} object with, amongst others, a \textit{country} field which is known to be required. In this case, $address \rightarrow country$ is technically an inter-parameter constraint. We choose to not include these as inter-parameter constraints, because of their frequent occurrence and triviality.


\subsection{Representation}
\label{sec:constraint_representation}

How constraints are represented can strongly impact the results. For example, if we choose to represent $A \rightarrow B \; \& \; C$ as $A \rightarrow B$ and $A \rightarrow C$, then we end up with twice the constraints. The same applies for $A \: || \: B \: \rightarrow C$. 

Generally, we opt to group logical ORs and logical ANDs together. This is done in order to match how constraints would be present in IF-statements; multiple conditions in the guard (left-side) would lead to a number of consequences in the body (right-side). If-statements are a particularly common control structure to encode constraints.



\subsection{Analysis}

We manually compared the output given by the approaches to the ground truth. If the identified constraint and ground truth constraint are logically equivalent, then we consider them to be the same constraint. Given that both approaches represent the output of constraints using logical formulations, this comparison can be done directly.

Sometimes only part of a constraint was identified. For example, given $A \rightarrow B \: \& \: C$, it would only identify $A \rightarrow C$. In these cases we deviate from the representation standard established in Section \ref{sec:constraint_representation}, and represent the constraint $A \rightarrow B$ as unidentified and $A \rightarrow C$ as identified.




\section{Results}
\subsection{RQ1: \rqone}

\subsubsection{Inter-parameter Constraints}
\label{sec:rq1-a}

We show the number of inter-parameter constraints identified by each approach in Table~\ref{table:results_coverage}.

\begin{table}
\centering
\caption{The total number of manually identified inter-parameter constraints, the number of constraints identified by the code and documentation analysis, with their respective false positives (FP), and the number of constraints that were identified by both.}
\label{table:results_coverage}
\begin{tabular}{lr|rr|rr|r}
\toprule
\textbf{}                         & \textbf{Total} & \textbf{Code} & \textbf{FP} & \textbf{Doc} & \textbf{FP} & \textbf{Both} \\ \midrule
\textbf{/payments}               			& 17             & 11          &   2          & 0     & 0	  & 0             \\ 
\textbf{/authorise}               			& 15             & 11          &   4 	      & 3     & 0	   & 2             \\ 
\textbf{/capture}                 			& 5              & 2             &   0	     & 1      & 0	  & 0             \\ 
\textbf{/storeDetail...} 	& 5              & 2             &   0         & 1        & 0	 & 1             \\ 
\textbf{/createAccount...}     	& 4              & 0             &   0         & 3        & 0	 & 0             \\ 
\textbf{/getAccount...}        	& 1              & 0             &   0         & 1        & 0	 & 0             \\ 
\textbf{/updateAccount...}     & 1              & 1             &    0        & 0         & 0	 & 0             \\ 
\textbf{/createAccount}           		& 1              & 0             &   1        & 1          & 0	 & 0             \\ 
\textbf{/uploadDocument}          	& 3              & 1             &   1      	 & 1         & 0	 & 1             \\ 
\textbf{/getCostEstimate}         		& 1              & 0             &   0    	 & 1         & 0	 & 0             \\ \midrule
\textbf{Total}         						& 53              & 28         &    8	    	 & 12       & 0	 & 4             \\ \bottomrule
\end{tabular}
\end{table}

We observe that code and documentation analysis together detected 36 ((28+12)-4) out of the 53 constraints. In other words, 68\% of the inter-parameter constraints.
We note that the approach was able to identify constraints in all APIs.

We also note that code and documentation analysis detect different constraints.
Between the 28 and 12 constraints found in the code and documentation approach, respectively, only four were found by both approaches. This indicates that both approaches are indeed complementary and, when used together, improve the overall results.

Finally, we also observe that code analysis detects more constraints, but with more false positives. The documentation analysis found fewer constraints, but did not produce any false positives. The code analysis approach detected around 2.5 times more constraints than documentation analysis.

\medskip
\subsubsection{Single-Parameter Constraints}
\label{sec:rq1-b}

We show the results in Table~\ref{table:results_code_pc}. Note that this only includes code analysis, since our documentation analysis approach is not set up to find single-parameter constraints. 

\begin{table}
\centering
\caption{The total amount of parameter constraints and the number of identified single-parameter constraints using code analysis.}
\label{table:results_code_pc}
\begin{tabular}{lrr} 
\toprule
                    & \textbf{Total} & \textbf{Identified} \\ \midrule
\textbf{/payments}  						& 9         & 8             \\
\textbf{/authorise} 						& 14         & 10             \\ 
\textbf{/capture}       					&     5      &       5           \\ 
\textbf{/storeDetailAndSubmit...}  &     4     &       4            \\ 
\textbf{/createAccountHolder}      &     4     &           1        \\ 
\textbf{/createAccount}     		    &      1     &          1    \\ \midrule
\textbf{Total:}      					    &      37     &          29    \\ \bottomrule
\end{tabular}
\end{table}

The code analysis approach detected 29 out of 37 the single-parameter constraints (or 78\%). For some endpoints, it manages to find all parameter constraints. 
The success of the approach in detecting single-parameter constraints since the code structures that handle these constraints are often simple to be parsed.
For example, the \emph{fraudOffset} parameter having to be smaller than 999 would be done with a check similar to \textit{if(request.getFraudOffset() < 999)}. Some notable challenging cases, which our approach failed, include regex patterns not being parsed to something meaningful, and the parsing of dates.

\begin{framed}
\noindent
The code and documentation approaches, when combined, identify 68\% of the inter-parameter constraints. The code analysis approach identifies 78\% of the single-parameter constraints.
\end{framed}

\subsection{RQ2: \rqtwo}

\subsubsection{Documentation Analysis}

We identify four reasons (lack of information, implicit references, value not detected, and unobserved constraints) that explain the failures in the documentation analysis approach. In Table~\ref{tab:rq2-doc}, we show how often each of them occurred.

\vspace{2mm}
\textbf{Lack of Information (A1).}
The most common reason for not identifying a constraint is the absence of information about the constraint in the documentation. For example, the two parameters \textit{paymentMethod.type} and \textit{returnUrl} in the /payments endpoint have a constraint between them, but the documentation does not mention it.

\vspace{2mm}
\textbf{Implicit References (A2).}
Some constraints were not detected due to the use of implicit information. 
There were a number of cases in which the OAS did include documentation on a constraint, but the description did not explicitly mention the name of a parameter (or at least not in a way that our approach could identify). For example, the description of the \textit{stateOrProvince} parameter states 'Required for the US and Canada'. Any human would know that the \textit{country} parameter's value being equal to 'US' or 'CA' would require \textit{stateOrProvince}.

\vspace{2mm}
\textbf{Value not Detected (A3).}
Finding the exact values that the constraints depend on can be difficult. As examples, for constraint $recurring.contract = "ONECLICK" $ $\rightarrow$ $ card.cvc$ the value \textit{ONECLICK} was not detected, and for the $country = "US" \rightarrow stateOrProvince$ constraint, the value \textit{US} was not detected. Although these values were present in the documentation, they did not have any special formatting, nor were they in the OpenAPI specifications, which makes them hard to be detected.



\vspace{2mm}
\textbf{Unobserved Constraints (A4).} 
Constraints may only be partially detected, which results in that constraint not being detected at all. For example, assume a constraint $A \rightarrow B \: \& \: C$. In the first step of documentation analysis (candidates inference), we only find that $A$ and $B$ co-occur. In the second step (validation of the candidates), we find that combinations including $A$ always fail, because both $B$ and $C$ are required. As a consequence, we fail to detect the actual constraint between $A$ and $B$.



\begin{table}
\centering
\caption{The reasons the documentation analysis was not able to identify an inter-parameter constraint. One constraint may not have been identified for multiple reasons.}
\label{tab:rq2-doc}
\begin{tabular}{lrrrr} 
\toprule
& \textbf{A1} & \textbf{A2} & \textbf{A3} & \textbf{A4} \\
\midrule
\textbf{/payments}  						& 15       &    2        	& 0                & 0                   \\ 
\textbf{/authorise}				 		& 7         &    2        	& 3               	& 2                    \\ 
\textbf{/capture}    						&   4     	 &    0        	&      0           &         0        \\ 
\textbf{/storeDetail...}  &     3   	&     1       	&     0             &      0        \\ 
\textbf{/createAccount...}      &      0    &     1      	&      0       		&      0           \\ 
\textbf{/getAccount...}       	&    0      &      0      &   0        	    &   0            \\ 
\textbf{/updateAcount...}    &      0    &      1     	&      0            &       0       \\ 
\textbf{/createAccount}       			&      0    &      0     	&      0            &       0      \\ 
\textbf{/uploadDocument}       		&      2    &      0     	&      0            &       0       \\ 
\textbf{/getCostEstimate}       		&     0     &      0     	&      0         &       0     \\ \midrule
\textbf{Total}       						&    30   &      7	    &     3          &       2       \\ \bottomrule
\end{tabular}
\end{table}

\begin{framed}
\noindent
The most common reason for the documentation approach to not detecting constraints is the absence of information that explicitly describes the constraints in the documentation. 
\end{framed}

\subsubsection{Code Analysis}

We identified eight reasons (parameter not detected, parameter de-referenced, static variable stack, pre-conditions, control structure, data flow, arithmetic constraint syntax, and framework) that make the code analysis to fail. In Table~\ref{tab:rq2-code}, we show their prevalence.

\vspace{2mm}
\textbf{Parameter not Detected (B1).}
We only identify constraints for the variables marked as parameters of the API. The list of parameters comes from the OpenAPI documentation. We have observed different reasons for a parameter not being documented: from a business perspective, these parameters may not be referenced because a function is in the process of being deprecated, or because certain functionality is only intended to be used by a select group of API consumers; from an API design perspective, these parameters might be missing because documentation still needs to be added.

As a possible solution for this problem, future implementations may consider all the fields from data model classes as parameters of the API. Given that the list of data model classes can be often inferred automatically (e.g., they all exist in the same package, or follow some naming convention), we see this as a viable alternative in cases where even the parameters that the API receives is not completely documented.

\vspace{2mm}
\textbf{Parameter De-referenced (B2).}
A large number of parameter references are de-referenced at some point.
Typically, we lose the reference to the parameter due to our parser not being able to parse all Java expressions. In this case study, this usually happened when new objects holding the values of parameters were created, either via traditional class instantiation or deserialisation.

For cases where new data objects are instantiated, and the parameter values are passed, we see a solution by improving the way our parser works. More specifically, references could be retained if we keep track of all the instantiated objects in our static variable stack. When parsing the construction of an object, we can theoretically keep track of which fields of that object correspond to which parameter(s). Due to time restrictions, we have not explored this in more detail.

However, we note that, in case of deserialisation, maintaining the references forms a significant challenge with no easy solution. This is mostly because the deserialisation steps are tightly interwoven with the framework that is being used at the company. 

\vspace{2mm}
\textbf{Static Variable Stack (B3).} 
Maintaining basic values in the variable stack is sufficient for inferring most inter-parameter constraints.
These basic values include Java primitives, strings, and parameter references. This tends to be enough since values in API requests are typically basic types. For example, a 'name' would be a string, and an 'amount' would be an integer. However, just maintaining basic values is not always enough, as often these parameter values as stored in more complex types (e.g., a domain object). Keeping these objects in the stack presents, however, an engineering challenge.

\vspace{2mm}
\textbf{Pre-conditions (B4).}
Some constraints have complex pre-conditions resulting in another parameter being required or not. For example, consider a hypothetical function $isValidIban(iban)$ in which validity of the IBAN itself depends on a large number of conditions, which the approach aims to parse. Typically, such preconditions exhibit expressions that are difficult for any static code analysis to parse. As a result, our approach would produce pre-conditions containing an often large number of unparsed elements. While somewhat related to B2 and B3, we put it as a separate challenge category as we conjecture this also requires a different solution.

\vspace{2mm}
\textbf{Control structure (B5).}
We observe that for-loops are used in a number of cases to validate parameter constraints within the source code itself. When for-loops were used, this was typically done for parameters with array values, e.g., ``people'': $[\{"name":"Frank", ...\}, ...]$. At times, it was also used for arithmetic constraints of the type $P1 + P2 + ... \; = \; Pn$. As discussed in Section \ref{sec:control_flow_graph}, evaluating the body of the for-loop once would be sufficient to infer the constraints. However, our for-loop parsing approach was limited in some cases. For example, when iterating a collection: $for (person : request.getPeople())$ or $for (i=0; i<people.size; i++)$. 

Theoretically, a more advanced analysis of for-loops could be needed. Consider a for-loop which iterates over a number of parameter objects. Each of these parameter objects could implement their own \textit{validate()} method. The aforementioned heuristics would not suffice. However, although not unlikely, we did not encounter such a structure within the code base.



%


\vspace{2mm}
\textbf{Data Flow (B6).}
In some cases, the limitations related to the tool's ability of tracking the data flow caused our approach to not identify the constraint. These limitations involved the lack of path sensitivity and not keeping track of fall-through conditions on certain branches.

Making static analysis path-sensitive is possible~\cite{dillig2008sound}. However, we currently do not see an easy way to implement it in our current approach. As such, within the analysis of parameter constraints, this is still an open challenge.

\vspace{2mm}
\textbf{Arithmetic Constraint Syntax (B7).}
Arithmetic constraints include parameters that are related to each other by means of arithmetic. For example, $A + B > 10$. 
Within our case study, such constraints were often directly present in the code, e.g., $A >= B$ would have a corresponding $if(A >= B)$ statement. However, other constraints were encoded through loops, e.g., $sum(split.value) = total$, which our tool was not able to properly detect. 

Arithmetic constraints can get as complex as mathematics itself. However, we observe in our case study that complex constraints are, in fact, rare. Common arithmetic constraints can be supported easily, whereas slightly more complicated constraints ($sum(splits) = total$) provide more challenges to overcome.

\vspace{2mm}
\textbf{Frameworks/Libraries (B8).}
The remote procedure call (RPC) based framework within Adyen was occasionally used to dynamically add new tasks. These tasks get passed through the framework, to then be handled as a kind of internal API request. Due to certain characteristics of the framework, such as multiple layers of abstraction, resolving which tasks get executed is especially difficult. 

\begin{table}
\centering
\caption{The reasons the code analysis was not able to identify an inter-parameter constraint. One constraint may not have been identified for multiple reasons.}
\label{tab:rq2-code}
\begin{tabular}{lrrrrrrrr} 
\toprule

& \textbf{B1} & \textbf{B2} & \textbf{B3} & \textbf{B4} & \textbf{B5} & \textbf{B6} & \textbf{B7} & \textbf{B8} \\
\midrule
\textbf{/payments} 						& 1         & 1          & 0          & 1         & 1 		 	 & 1     & 0      & 2 \\ 
\textbf{/authorise} 						& 0         & 3          & 2          & 1         & 3 		 	 & 0     & 1      & 0 \\ 
\textbf{/capture}       					& 0         & 3          & 2          & 0         & 3 		 	 & 0     & 1      & 0 \\ 
\textbf{/storeDetail...}  & 0         & 0          & 0          & 0         & 0 		 	 & 3     & 0      & 3 \\ 
\textbf{/createAccount...}      & 0         & 2          & 2          & 0         & 4 		 	 & 2     & 0      & 0 \\ 
\textbf{/getAccount...}       	& 0         & 0          & 0          & 0         & 0		 	 & 1     & 0      & 0 \\ 
\textbf{/updateAccount...}  & 0         & 0          & 0          & 0         & 0 		 	 & 0     & 0      & 0 \\ 
\textbf{/createAccount}       		    & 0         & 1          & 0          & 0         & 0 		 	 & 0     & 0      & 0 \\ 
\textbf{/uploadDocument}       	    & 0         & 2          & 0          & 0         & 0 		 	 & 0     & 0      & 0 \\ 
\textbf{/getCostEstimate}     	    & 0         & 1          & 1          & 0        & 0 		 	 & 0     & 0      & 0 \\ 
\midrule
\textbf{Total}       					    & 1         & 13        & 7          & 2        & 11 		 	 & 7     & 2      & 5 \\ 

\bottomrule

\end{tabular}
\end{table}

\begin{framed}
\noindent 
The challenges in the code analysis are the engineering of a sound static code analyser that is sensitive to data-flow, that maintains longer parameter references throughout the API's code, and that is able to symbolically execute the several libraries and frameworks used by the API.
\end{framed}

\section{Discussion}
\label{cha:discussion}
In this section, we compare our work to the existing literature, how much we expect our results to generalize to other APIs and, finally, the threats to the validity of our wok.

\subsection{Comparison to Previous Works}

We argue that the major difference between our work and the related work was our need to work with a large and complex API.
Existing works that specifically focused on inter-parameter constraints for Web APIs evaluated APIs with a small number of parameters (i.e., around 5 parameters). The works by Gao et al.~\cite{gao2014inferring} and Wu et al.~\cite{wu2013inferring} had an overall recall of 95.5\%. We obtain a recall of 23\% using a documentation analysis approach; with a more extensive code analysis approach we obtain a recall of 53\%. We argue that this is largely due to the differences in the complexities of the studied APIs.

Concerning code analysis, Wu et al.~\cite{wu2013inferring} is, to the best of our knowledge, the only work which describes the analysis of code to infer inter-parameter constraints. We highlight several differences between our works. First and foremost, their approach performs data-flow analysis, whereas we extract a constraint structure directly. Secondly, their approach does not address some of the features that exist in Adyen's code base. Key examples are the analysis of methods separately, as opposed to within its given context, and the lack of support for value-dependent constraints. The latter effectively made their approach to not support single-parameter constraints or any constraints such as $X \:= \:V \rightarrow Y$.

\subsection{Generalizability of Our Findings}

We conjecture that software development teams developing web in any programming language rely on similar sets of (language) expressions and make use of similar frameworks in the process. To this extent, we conjecture that a large number of challenges we describe above would apply to other software companies, even those using a different programming language.

The studied API makes use of RPC. Remote procedure call (RPC) based frameworks treat API requests as calls to functions, where the arguments to the function are put in either the query string or the body. This is contrary to REST, which often involves path parameters such as $/store/orders/{orderID}$. We nevertheless do not see a reason to believe that our approach would be dependent of how the API is published.

The size and complexity of the API may have influence on the timeliness of the feedback.
Complex APIs, such as ours, are likely to have more code, making fast sound analysis a point of focus. After all, extensively analyzing every method might be too computationally intensive. We note that the approach we describe in this paper is able to analyze methods quickly enough for large software systems. However, the challenges we identified seem to require more sophisticated static analysis techniques. We argue that future researchers should monitor the trade-offs between a more accurate approach versus the time it will take to run.

\subsection{Security Concerns}

Constraints can reveal information companies do not want the public to know. After all, the constraints that are extracted from the source code reflect the code to a certain extent. 

Attackers might make use of this information.
For example, the constraints provide information on boundary conditions used within the code. A malicious third party could potentially use this information to exploit the API more easily. As such, automatically making all identified constraints available to the general public is not preferable. Moreover, the constraints may reveal features that are still in development or that are only intended to be used by a select group of API consumers.

Therefore, we argue that the output of such tools should be carefully analyzed by developers and security experts before becoming public.

\subsection{Threats to Validity}

In the following, we discuss the threats to the validity of this paper and actions we took to mitigate them.

\subsubsection{Internal Validity} The ground truth we used consists out of a representative set of constraints which we manually collected for each of the selected endpoints. In this process, we thoroughly inspected the related code and verified them by making requests to the API. As such, we are confident that we have a correct set of constraints. When selecting the set of APIs to study, we made sure to select a diverse set of APIs and constraints. As such, we are also confident that the results represent the different APIs within Adyen well.

\subsubsection{External Validity} Given that this research is a case study done in one company, research into other complex APIs is needed for further generalization of the results. However, given the size and scale of Adyen's software, we are confident that the results found in our study can be representative for other large-scale APIs.

\section{Conclusion}
\label{cha:conc}

Understanding the parameter constraints of Web APIs is fundamental when comes to their usability. On one hand, documenting them all may require a large effort from the development and documentation teams; on the other hand, incomplete or incorrect documentation on these constraints can waste a lot of time and cause costly integration faults. 

In this paper, we describe our case study at Adyen, where we experiment the effectiveness of the existing approaches for inferring parameter constraints in our Web APIs. 
Our results show that the documentation and code-based approaches can, together, identify a total of 68\% of the inter-parameter constraints in the large and complex APIs we use as case study. 

While we believe that the current results are already promising and development teams can use such approaches to support their documentation teams, there is still much room for improvement. We hope that the list of challenges we discuss in this paper will pave the road for future research on the topic.

\renewcommand*{\bibfont}{\footnotesize}
\bibliographystyle{IEEEtranN}
\bibliography{paper}

\end{document}